\documentclass[floats,aps,amssymb,nofootinbib]{revtex4}
\usepackage{amssymb}
\usepackage{graphics}
\usepackage{amsmath}
\usepackage{amsfonts}
\usepackage{bm}
\usepackage[]{latexsym}
\usepackage{float}
\usepackage{graphicx}
\usepackage{hyperref}
\newcommand{\ket}[1]{\left| #1 \right\rangle}

\newcommand{\inn}[2]{\left\langle #1 \middle| #2 \right\rangle}

\begin{document}
\title{Chiral oscillations in the non-relativistic regime}

\author{Victor A. S. V. Bittencourt\footnote{victor.bittencourt@mpl.mpg.de},$^{\hspace{0.3mm}1}$Alex E. Bernardini\footnote{alexeb@ufscar.br},$^{\hspace{0.3mm}2}$Massimo Blasone\footnote{blasone@sa.infn.it}$^{\hspace{0.3mm}3,4}$} \affiliation
{$^1$Max Planck Institute for the Science of Light, Staudtstra\ss e 2, PLZ 91058, Erlangen, Germany. \\
$^2$Departamento de F\'isica, Universidade Federal de S\~ao Carlos, P.O. Box 676, 13565-905 S\~ao Carlos, S\~ao Paulo. \\
$^3$Dipartimento di Fisica, Universit\`a di Salerno, Via Giovanni Paolo II, 132 I-84084 Fisciano (SA), Italy.\\ $^4$INFN, Sezione di Napoli, Gruppo collegato di Salerno, Italy.}

\begin{abstract}
Massive Dirac particles are a superposition of left and right chiral components. Since chirality is not a conserved quantity, the free Dirac Hamiltonian evolution induces chiral quantum oscillations, a phenomenon related to the \textit{Zitterbewegung}, the trembling motion of free propagating particles. While not observable for particles in relativistic dynamical regimes, chiral oscillations become relevant when the particle's rest energy is comparable to its momentum. In this paper, we quantify the effect of chiral oscillations on the non-relativistic evolution of a particle state described as a Dirac bispinor and specialize our results to describe the interplay between chiral and flavor oscillations of non-relativistic neutrinos: we compute the time-averaged survival probability and observe an energy-dependent depletion of the quantity when compared to the standard oscillation formula. In the non-relativistic regime, this depletion due to chiral oscillations can be as large as 40$\%$. Finally, we discuss the relevance of chiral oscillations in upcoming experiments which will probe the cosmic neutrino background.
\end{abstract}

\maketitle

\vskip -1.0 truecm 

\section{Introduction}
\label{intro}
Dirac equation has unique dynamical predictions for fermionic particles, from the Klein paradox \cite{Klein:1929,Itzykson}, related to pair production in scattering problems, to the \textit{Zitterbewegung}, the trembling motion of free relativistic states \cite{Schrodinger:1930}. Formally, the solutions of the Dirac equation are given in terms of 4-component spinors, or Dirac bispinors. From  a group-theoretical perspective, those objects belong to the irreducible representations of the \textit{complete} Lorentz group and are constructed by combining Weyl spinors of different chiralities \cite{WuTung}. Any Dirac bispinor has left and right-chirality components which are dynamically coupled by the mass term of the Dirac equation. Since a bispinor with definite chirality cannot be an eigenvector of the massive Dirac Hamiltonian, the free evolution of such state will induce left-right \textit{chiral oscillations} \cite{DeLeo:1998}. The degree of left and right chirality superposition in a Dirac bispinor depends on the energy to mass ratio, thus  chiral oscillations are usually not relevant for the description of relativistic particles, but are prominent for dynamical regimes in which the momentum of the particle is comparable to (or smaller than) its mass.

The non-relativistic regime is also interesting for exploring the connection between the \textit{Zitterbewegung} effect and chiral oscillations. As discussed in \cite{Bernardini:2006epjc}, the trembling motion of free Dirac particles has an intimate relation to  chiral oscillations. Both are related to the fact that a state with initial definite chirality must be described as a superposition of positive and negative energy solutions of the Dirac equation. Although the \textit{Zitterbewegung} effect is associated with a fast frequency, which is averaged out in typical high energy systems, it has already been considered to explain the Darwin correction term in hydrogenionic atoms \cite{Itzykson} and it has been probed in Dirac-like systems, such as graphene \cite{CastroNeto:2009} and trapped ions  \cite{Lamata:2007,Lamata:2011}. For example, the graphene analogous to the \textit{Zitterbewegung} \cite{Rusin:2008} can be measured via laser excitations \cite{Rusin:2009}. While mono-layer graphene is described by the massless Dirac equation, bilayer graphene displays effects associated with the mass term of the Dirac equation \cite{CastroNeto:2009} plus an effective non-minimal coupling \cite{Bittencourt:2017}, and is a system in which chiral oscillations and their relation to the \textit{Zitterbewegung} could be probed. 

In this paper, we are concerned with the non-relativistic limit of chiral oscillations within the framework of the Dirac equation. Since the chiral oscillation amplitude depends on the mass to energy ratio \cite{Bernardini:2004epjc,Bernardini:2005prd}, the minimum survival probability of an initial state with definite chirality could be averaged out in the non-relativistic limit. Therefore, in the non-relativistic limit we specialize our discussion to (Dirac) neutrinos which indeed are ideal candidates for the study of chiral oscillations due to the chiral nature of weak interaction processes in which they are produced and/or detected \cite{Giunti}. The Cosmic Neutrino Background (C$\nu$B) is an important example of neutrinos in non-relativistic regime. The C$\nu$B is the neutrino counterpart of the Cosmic Microwave Background (CMB) for photons: it is composed by decoupled relic neutrinos evolving in an expanding background \cite{Giunti}. The  PTOLEMY project\cite{Baracchini:2018wwj} aims at detecting the C$\nu$B via neutrino capture on tritium, a goal which should be reached within the next years. In previous studies\cite{Long:2014zva,Roulet:2018fyh}, it has been noticed that the expected event rate of C$\nu$B capture on tritium will exhibit a depletion. We show that such a depletion for Dirac neutrinos can be understood as a manifestation of chiral oscillations and give a precise quantification of it. Furthermore, the capture rate depends on the nature of the neutrinos: Majorana or Dirac. While chiral oscillations are present in both cases, Majorana neutrinos are subjected to more measurable oscillation channels than the Dirac case, e.g. to right-chiral and positive helicity states.

Although the interplay between chiral and flavor oscillations is very small in the ultra-relativistic regime \cite{Bernardini:2004epjc,Bernardini:2005prd,Nishi:2006prd,Bernardini:2011fph}, it is relevant for describing dynamical features of non-relativistic neutrinos. We  describe chiral oscillations in the context of two-flavor neutrino propagation, including flavor oscillations. In particular, we compute the effects of chiral conversion on the averaged flavor oscillation, that is, the survival probability of a neutrino of a given flavor averaged over one period of flavor oscillation. The chiral oscillations corrections are relevant when the particle's momentum is comparable with the lightest neutrino mass and, in the non-relativistic regime, the maximum difference is 40$\%$, a prediction consistent with the preliminary discussion pointed in \cite{Roulet:2018fyh} in connection with C$\nu$B detection.  A more in depth discussion of chiral oscillations in the context of the C$\nu$B can be found in \cite{Ge:2020}. To highlight the main dynamical features associated to this phenomenon, we adopt a simple plane wave description and leave a more realistic  treatment involving wave packets \cite{Bernardini:2004epjc,Bernardini:2005prd,Bernardini:2011fph} for future work. Although we focus on Dirac neutrinos, we discuss chiral oscillations effects in Majorana neutrinos and provide a detailed calculation of  them for the general Majorana-Dirac mass term in the appendix.

\section{Chirality and chiral oscillations in bispinor dynamics}
\label{sec:1}

Through this paper we will describe dynamical features of free massive fermionic particles in the context of relativistic quantum mechanics. We thus consider the temporal evolution as given by the Dirac equation (hereafter we adopt natural units $\hbar = c = 1$)
\begin{equation}
\label{MainDiracEq}
\hat{H}_D \ket{\psi} = \left(\hat{\bold{p}} \cdot \hat{\bold{\alpha}} + m \hat{\beta}\right) \ket{\psi} = i \partial_t \ket{\psi},
\end{equation}
where $\hat{\bold{p}}$ is the momentum operator and $m$ is the particle's mass. The operators $\hat{\alpha}_i$ ($i=x,y,z$) and $\hat{\beta}$ are the $4 \times 4$ Dirac matrices satisfying the anti-commutation relations $ \hat{\alpha}_i \hat{\alpha}_j +\hat{\alpha}_i \hat{\alpha}_j = 2 \delta_{ij} \hat{I}_4$, $\hat{\alpha}_i \hat{\beta} +\hat{\alpha}_i \hat{\beta} = 0$, and $\hat{\beta}^2 = \hat{I}_4$. 

In the language of group theory, Dirac equation is the dynamical equation for the irreducible representations of the \textit{complete} Lorentz group\footnote{The complete Lorentz group is the proper Lorentz group plus Parity.}, which are the well-known Dirac bispinors \cite{WuTung}. These are constructed by means of the irreducible representations of the (proper) Lorentz group, the Weyl spinors. The later belong to the irreps of a group isomorphic to the $SU(2)$ and as such they carry an intrinsic degree of freedom - the spin. The elements of the two disconnected irreps of the proper Lorentz group are labeled as left and right-handed spinors. Since parity connects the left and right representations, the irreps of the complete Lorentz group are obtained by combining these representations, and thus the Dirac bispinors carry not only the spin, but also another intrinsic discrete degree of freedom, the chirality. Any massive Dirac bispinor is thus a combination of left and right-handed spinors. 

Turning our attention to the Dirac equation framework, chirality is the average value of the chiral operator $\hat{\gamma}_5 = -i
 \hat{\alpha}_x  \hat{\alpha}_y  \hat{\alpha}_z$. While such operator commutes with the momentum term of the Dirac Hamiltonian, it does not commute with the mass term and thus it is not a dynamically conserved quantity for bispinor states describing massive particles. This can be better appreciated by considering the chiral representation of the Dirac matrices, in which $\hat{\gamma}_5 = {\rm{diag}}\{\hat{I}_2,- \hat{I}_2\}$\footnote{In the chiral representation $\hat{\alpha} =\begin{bmatrix} \hat{\bold{\sigma}} & 0 \\ 0 & -\hat{\bold{\sigma}} \end{bmatrix} $ and $\hat{\beta} = \begin{bmatrix} 0 & \hat{I}_2 \\  \hat{I}_2 & 0 \end{bmatrix}$.}. Any bispinor $\ket{\xi}$ can be written in this representation as
\begin{equation}
\label{LRDecomp}
\ket{\xi} = \begin{bmatrix}
  \ket{\xi_R} \\
    \ket{\xi_L} 
\end{bmatrix},
\end{equation}
where $\ket{\xi_{R, L}}$ are, respective, the positive (right-handed) and negative (left-handed) chirality two-component spinors. The Dirac equation $\hat{H}_D \ket{\xi} = i \dot{\ket{ \xi}}$ can then be written as
\begin{equation}
\label{LREquations}
\begin{aligned}
\hat{\bold{p}}\cdot \hat{\bold{\sigma}} \ket{\xi_R} + m \ket{\xi_L} &= i \partial_t \ket{\xi_R}, \\
-\hat{\bold{p}}\cdot \hat{\bold{\sigma}} \ket{\xi_L} + m \ket{\xi_R} &=i \partial_t \ket{\xi_L},
\end{aligned}
\end{equation}
from which it  is clear that the mass term $m \hat{\beta}$ connects the left and right-handed components of the bispinor. 

For a given initial state, evolution under the free Dirac Hamiltonian $\hat{H}_D$ induces left-right chiral oscillations. In order to describe this dynamical effect we first introduce the positive and negative plane wave solutions of the Dirac equation, $\ket{\psi_+(\bold{x}, t)} = e^{i \bold{p} \cdot \bold{x} - i E_{p,m} t }\ket{u_s (\bold{p}, m)} $ and  $\ket{\psi_-(\bold{x}, t)} = e^{-i \bold{p} \cdot \bold{x} + i E_{p,m} t }\ket{v_s (\bold{p}, m)} $ which are given in terms of the bispinors
\begin{equation}
\ket{u_s (\bold{p}, m)} = \sqrt{\frac{E_{p,m} + m}{4 E_{p,m}}} \begin{bmatrix}
    \left(1 + \frac{\bold{p}\cdot \hat{\bold{\sigma}}}{E_{p,m} + m}\right) \ket{\eta_s(\bold{p})}  \\
   \left(1 - \frac{\bold{p}\cdot \hat{\bold{\sigma}}}{E_{p,m} + m}\right) \ket{\eta_s(\bold{p})}  
\end{bmatrix}, \hspace{0.5 cm} \ket{v_s (\bold{p}, m)} = \sqrt{\frac{E_{p,m} + m}{4 E_{p,m}}} \begin{bmatrix}
    \left(1 + \frac{\bold{p}\cdot \hat{\bold{\sigma}}}{E_{p,m} + m}\right) \ket{\eta_s(\bold{p})}  \\
   -\left(1 - \frac{\bold{p}\cdot \hat{\bold{\sigma}}}{E_{p,m} + m}\right) \ket{\eta_s(\bold{p})}  
\end{bmatrix},
\end{equation}
where $E_{p,m} = \sqrt{p^2 + m^2}$. In the above equations, $\ket{\eta_s(\bold{p})}$ is a two component spinor that depends on the spin polarization of the particle. We notice that the orthogonality relations read: $\inn{u_s (\bold{p}, m)}{v_s (-\bold{p}, m)}=\inn{v_s (-\bold{p}, m)}{u_s (\bold{p}, m)}=0$, $\inn{u_s (\bold{p}, m)}{u_s (\bold{p}, m)}=\inn{v_s (\bold{p}, m)}{v_s (\bold{p}, m)}=1$. For now on, we describe the solutions of the Dirac equation with helicity bispinors, that is, we assume that $\ket{\eta_s(\bold{p})}$ are eigenstates of the Helicity operator $\frac{\bold{p}\cdot \hat{\bold{\sigma}}}{p}$. This choice is convenient since helicity is a conserved quantity and all the relevant dynamical features will be entirely related to chiral oscillations. Moreover, we simplify our analysis by considering one-dimensional propagation along the $\bold{e}_z$ direction, such that
\begin{equation}
\label{DiracBispinors}
\ket{u_\pm (p, m)} = \sqrt{\frac{E_{p,m} + m}{4 E_{p,m}}} \begin{bmatrix}
    \left(1 \pm \frac{p}{E_{p,m} + m}\right) \ket{\pm}  \\
   \left(1 \mp \frac{p}{E_{p,m} + m}\right) \ket{\pm}  
\end{bmatrix}, \hspace{0.5 cm} \ket{v_\pm (p, m)} = \sqrt{\frac{E_{p,m} + m}{4 E_{p,m}}} \begin{bmatrix}
    \left(1 \pm \frac{p}{E_{p,m} + m}\right) \ket{\pm}  \\
   -\left(1\mp \frac{p}{E_{p,m} + m}\right) \ket{\pm}  
\end{bmatrix},
\end{equation}
where $\ket{\pm}$ are the eigenstates of the Pauli matrix $\hat{\sigma}_z$. For ultra-relativistic particles, $m/p\rightarrow 0$ and positive helicity bispinors have only right-handed components while negative helicity bispinors have only left-handed components. On the other hand, for $p/m \rightarrow 0$ the left and right-handed components are equal irrespective of the helicity.

We now describe chiral oscillations by considering the temporal evolution of the initial state $\ket{\psi(0)}  = [0, \, 0, \, 0,\, 1]^T$ which has negative helicity and negative chirality: $\hat{\gamma}_5 \ket{\psi (0)} = -\ket{\psi (0)}$. As we are dealing with plane wave states propagating one-dimensionally, we consider the dynamical evolution in momentum space \cite{Bernardini:2005prd}. The time evolved state $\ket{\psi_m(t)}$ is given by
\begin{equation}
\label{TempEvo}
\begin{aligned}
\ket{\psi_m(t)} &= e^{-i \hat{H}_D t} \ket{\psi(0)} \\
&= \sqrt{\frac{E_{p,m} + m}{4 E_{p,m}}}\left[ \left( 1 + \frac{p}{E_{p,m} + m} \right) e^{- i E_{p,m} t }\ket{u_- (p,m)} -  \left( 1 - \frac{p}{E_{p,m} + m} \right) e^{i E_{p,m} t} \ket{v_- (-p,m)} \right], 
\end{aligned}
\end{equation}
Chiral oscillations are generated by the massive character of the particle and by the fact that the initial state is a superposition of positive and negative energy eigenstates of the Dirac Hamiltonian. The later is also responsible for the relation between chiral oscillations and the \textit{Zitterbewegung} effect \cite{Bernardini:2006epjc}. The survival probability of the initial state $\mathcal{P}(t)$ is given by
\begin{equation}
\label{chiralosc01}
\mathcal{P}(t) = \vert \inn{\psi_m(0)}{ \psi_m(t) }\vert^2 = 1 - \frac{m^2}{E_{p,m} ^2} \sin^2\left( E_{p,m} t\right),
\end{equation}
while the (transition) probability of being in a positive chirality state is given by $\mathcal{P}_T(t)= 1-\mathcal{P}(t) $,
and the average value of the chiral operator $\langle \hat{\gamma}_5 \rangle (t)$ reads
\begin{equation}
\label{avgchir}
\langle \hat{\gamma}_5\rangle (t) = \langle \psi_m(t)\vert \hat{\gamma}_5 \ket{\psi_m(t)} = -1 + \frac{2 m^2}{E_{p,m} ^2} \sin^2\left( E_{p,m} t\right).
\end{equation}

\begin{figure}[t]
\centering
\includegraphics[width = 14 cm]{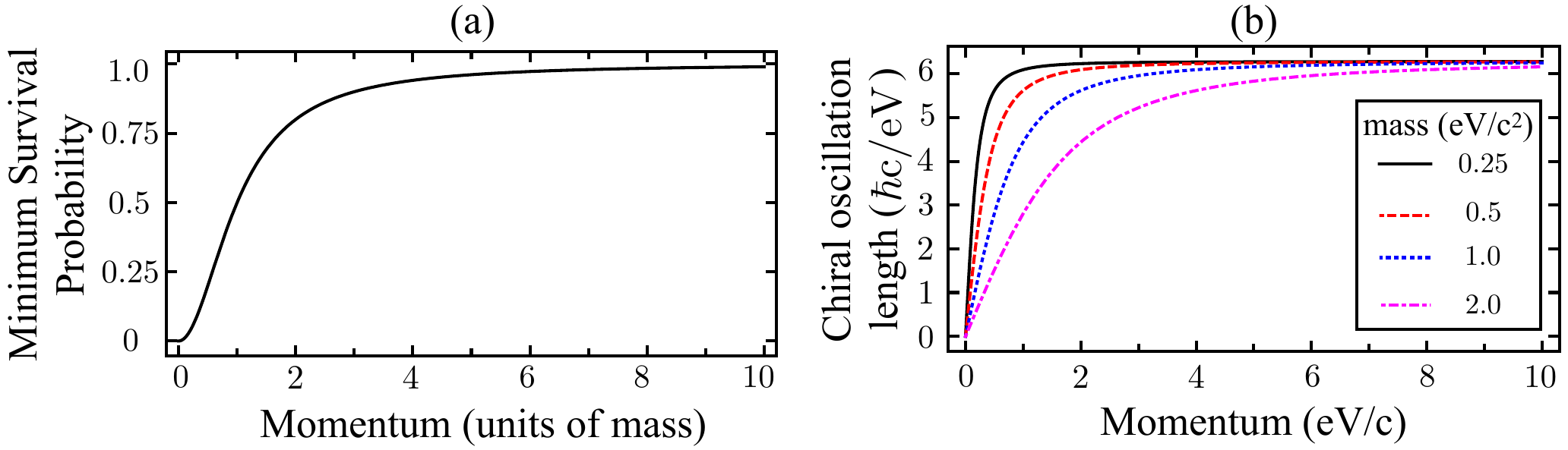}
\caption{(a) Minimum survival probability as a function of the momentum $p$ in units of mass; (b) Chiral oscillation length as a function of the momentum for different values of the mass.}
\label{Fig:01}
\end{figure}

According to \eqref{chiralosc01} the minimum survival probability is $\mathcal{P}_{\rm{min}}=1- \frac{m^2}{E_{p,m}^2}$, and the frequency of the chiral oscillations is (in natural units) $E_{p,m}$. Therefore, the corresponding period of one chiral oscillation is $\tau_{{\rm{Ch}}} = 2 \pi/E_{p,m}$ and as the particle propagates freely, the length corresponding to one chiral oscillation can be evaluated as $l_{\rm{Ch}} = p \tau / E_{p,m} = 2 \pi \frac{p}{E_{p,m}^2}$. In Fig.~\ref{Fig:01} we show $\mathcal{P}_{\rm{min}}$ as a function of $p/m$ (a) and  $l_{\rm{Ch}}$ as a function of $p$ for several mass values (b). For non-relativistic states, i.e. states for which $p \sim m$, chiral oscillations play an important role and affect significantly the probability of the state to be in its initial configuration. In fact, for $p \ll m$ the minimum survival probability is $\mathcal{P}_{\rm{min}} \sim \frac{p^2}{m^2}$ which vanishes as $p\rightarrow 0$. In this case, since the eigenspinors are a maximal superposition of left and right chiralities, the free evolution induces a complete oscillation from left to right chiral components. This can also be seen in \eqref{avgchir}: for $m \sim E_{p,m}$ the oscillations of the average chirality have the maximal amplitude between the initial value $-1$ and $1$. For $p \gg m$, chiral oscillations are less relevant for the state dynamics. In the limit $p\rightarrow \infty$ chirality and helicity coincide, thus the eigenspinors \eqref{DiracBispinors} have definite chirality. In fact \eqref{avgchir} is constant in the later limit. The chiral oscillation length $l_{\rm{Ch}}$ is shown in Fig.~\ref{Fig:01} (b) for several values of the mass. For masses in the range of ${\rm{eV}}/c^2$, the expected chiral oscillation length is of the order of $1 \hbar c/ {\rm{eV}} \sim 10^{-6}$ m. 

Our analysis has focused on Dirac particles with states given in terms of Dirac bispinors. For Majorana particles, one should have in mind that the mass term is proportional to the charge-conjugated spinor. In fact, the Dirac equation with the Majorana mass term, which we call Majorana equation, reads \cite{Pal:2011,Dvornikov:2009,Arodz2:2019,Hashimi:2017}
\begin{equation}
i \partial \ket{\psi} = \hat{\bold{p}} \cdot \hat{\bold{\alpha}} \ket{\psi} + m \hat{\beta} \ket{\psi}^c,
\end{equation}
where $\ket{\psi}^c = i \hat{\beta} \hat{\alpha}_y \ket{\psi}^*$ \cite{Cheng} is the charge conjugated bispinor. If the bispinor $\ket{\psi}$ is its own charge conjugated, that is, if $\ket{\psi}^c = \ket{\psi}$, then its temporal evolution is given by the ``usual'' Dirac equation \eqref{MainDiracEq}. This last condition is known as the Majorana condition. It can be shown (see the Appendix) that for any bispinor satisfying the Majorana condition $\langle \hat{\gamma}_5 \rangle = 0$. In other words, any bispinor that is its own self conjugate has zero average chirality. Since the Majorana condition is preserved under the time evolution, there is no chiral oscillations in bispinors satisfying the Majorana condition, irrespective if the bispinor is an eigenstate of the Hamiltonian. Notice that this is in stark contrast with the Dirac bispinor case: a Dirac bispinor is an unconstrained object whose time evolution is given by the usual Dirac equation. The average chirality of a Dirac bispinor is only constant if it is an eigenstate of the Dirac Hamiltonian.

Nevertheless, given the intrinsic handedness of the weak interaction, we can consider a bispinor that is initially in a left handed and negative helicity state but whose time evolution is given by the Majorana equation. In this case, we follow the formalism of \cite{Esposito:1997,Esposito:1998} and obtain that the survival probability of such state is the same as the one given in \eqref{chiralosc01} while there is a transition probability to a right-handed component, associated with the charge conjugation of the left-handed initial state. This fact was also briefly quoted in \cite{Ge:2020}. A more general situation includes both Dirac and Majorana mass terms with two non-degenerate Majorana masses, which we describe in the appendix.

\section{Flavor mixing and chiral oscillations in non-relativistic regime}
\label{sec:02}

We now study the effects of chiral oscillations in non-relativistic neutrino mixing. The state of a neutrino of flavor $\alpha$ at a given $t$ is given by the superposition of mass states:
\begin{equation}
\label{Flavor01}
\ket{\nu_\alpha (t)} = \sum_{i} U_{\alpha,i} \ket{\psi_{m_i} (t)} \otimes \ket{\nu_i},
\end{equation}
where $U$ is the mixing matrix, and $\ket{\psi_{m_i} (t)}$ are the bispinors describing the temporal evolution of the mass eigenstate $\ket{\nu_i}$ with mass $m_i$ \cite{Bernardini:2004epjc,Bernardini:2005prd}. The state at $t=0$ reads
\begin{equation}
\label{Flavor02}
\ket{\nu_\alpha (0)} =\ket{\psi (0)}  \otimes  \sum_{i} U_{\alpha,i}\ket{\nu_i} = \ket{\psi (0)} \otimes \ket{\nu_\alpha},
\end{equation}
with $\ket{\nu_\alpha} \equiv  \sum_{i} U_{\alpha,i}\ket{\nu_i}$, and thus $ \ket{\psi_{m_i} (t=0)} = \ket{\psi (0)} $. Since weak interaction processes only create left handed neutrinos, for now own, we take $\ket{\psi (0)} $ as the left handed bispinor with negative helicity of the last section. The temporal evolved flavor state is therefore
\begin{equation}
\ket{\nu_\alpha (t)}= \sum_{i} \sum_{\beta }U_{\alpha,i} U_{\beta,i}^* \ket{\psi_{m_i} (t)} \otimes \ket{\nu_\beta},
\end{equation}
where the $\ket{\psi_{m_i} (t)}$ are given by eq.~\eqref{TempEvo} with the substitution $m \rightarrow m_i$. The survival probability, i.e. the probability of the state $\ket{\nu_\alpha (t)}$ to be a left-handed state of $\alpha$ flavor, reads
\begin{equation}
\mathcal{P}_{\alpha \rightarrow \alpha}= \vert \inn{\nu_\alpha (0)}{\nu_\alpha (t)} \vert^2= \vert \sum_{i} \vert U_{\alpha, i} \vert^2 \inn{\psi(0)}{\psi_{m_i}(t)} \vert^2.
\end{equation}

For two flavors mixing, the time evolution of an initial electron neutrino state is \cite{Bernardini:2005prd}
\begin{equation}
\ket{\nu_e (t)} = \left[ \cos^2(\theta) \ket{\psi_{m_1} (t)} + \sin^2(\theta) \ket{\psi_{m_2} (t)} \right]\otimes \ket{\nu_e} + \left[  \ket{\psi_{m_1} (t)}- \ket{\psi_{m_2} (t)} \right] \sin(\theta) \cos(\theta) \ket{\nu_\mu},
\end{equation}
and the survival probability can be decomposed as
\begin{equation}
\label{survivProbE}
\mathcal{P}_{e\rightarrow e}(t) = \mathcal{P}_{e \rightarrow e}^{S}(t) + \mathcal{A}_e + \mathcal{B}_e.
\end{equation}
In this formula $\mathcal{P}_{e \rightarrow e}^{S}(t)$ is the standard flavor oscillation formula
\begin{equation}
\label{standard}
\mathcal{P}_{e \rightarrow e}^{S}(t) = 1 - \sin^2(2 \theta) \sin^2\left( \frac{E_{p, m_2} - E_{p, m_1}}{2} t \right)
\end{equation}
and
\begin{equation}
\label{AddFacs}
\begin{aligned}
\mathcal{A}_e(t) &= - \left[\frac{m_1}{E_{p,m_1}} \cos^2(\theta) \sin \left( E_{p,m_1} t \right)+ \frac{m_2}{E_{p,m_2}} \sin^2(\theta) \sin \left( E_{p,m_2} t \right) \right]^2, \\
\mathcal{B}_e(t) &= \frac{1}{2} \sin^2(2 \theta) \sin(E_{p,m_1} t) \sin(E_{p,m_2} t)\left(\frac{p^2 +m_1 m_2}{E_{p,m_1} E_{p,m_2}} -1\right),
\end{aligned}
\end{equation}
are correction terms due to the bispinor structure. Those corrections include an interplay between chiral and flavor oscillations effects and are in agreement with results presented in the literature \cite{Nishi:2006prd}. 

While the standard flavor oscillations have a time scale set by the energy difference $E_{p, m_2} - E_{p, m_1}$, chiral oscillations depend roughly on the mass-momentum ratio (see the previous section). The terms $\mathcal{A}_e$ and $\mathcal{B}_e$ depend non-trivially both on the energy difference and on the individual energies $E_{p, m_2}$ and $E_{p, m_1}$. In the limit $p \ll m_{1,2}$, $E_{p, m_2} - E_{p, m_1} \sim m_2 - m_1 + \mathcal{O}[p^2/m]$ and if the mass difference is $\vert m_2 - m_1 \vert \gg m_{2,1}$ or $\ll m_{2,1}$, there will be no interference effects between flavor and the chiral oscillations. Those phenomena can be individually identified in the survival probability. If $\vert m_2 - m_1 \vert$ is comparable to one of the masses, there will be an interference between flavor and chiral oscillations. We furthermore notice that the term $\mathcal{B}_e(t)$ is equivalent to corrections obtained via a quantum field description of flavor oscillations \cite{Blasone:1995aop,Blasone:1998hf}. In fact, after some algebra we obtain that $ \mathcal{P}_{e \rightarrow e}^{S}(t) + \mathcal{B}_e(t)$ reproduces eq. (30) of \cite{Blasone:1998hf} (see also \cite{Blasone:2019}). In the quantum field framework, the corrections have a similar origin to the usual Dirac \textit{Zitterbewegung}, being originated from the fact that the flavor ladder operator has contributions from both particle and antiparticle operators of the massive fields. Since those corrections do not take into account chiral oscillations, we interpret $\mathcal{A}_e(t)$ as the corrections due to chiral oscillations.

\begin{figure}[t]
\centering
\includegraphics[width = 15 cm]{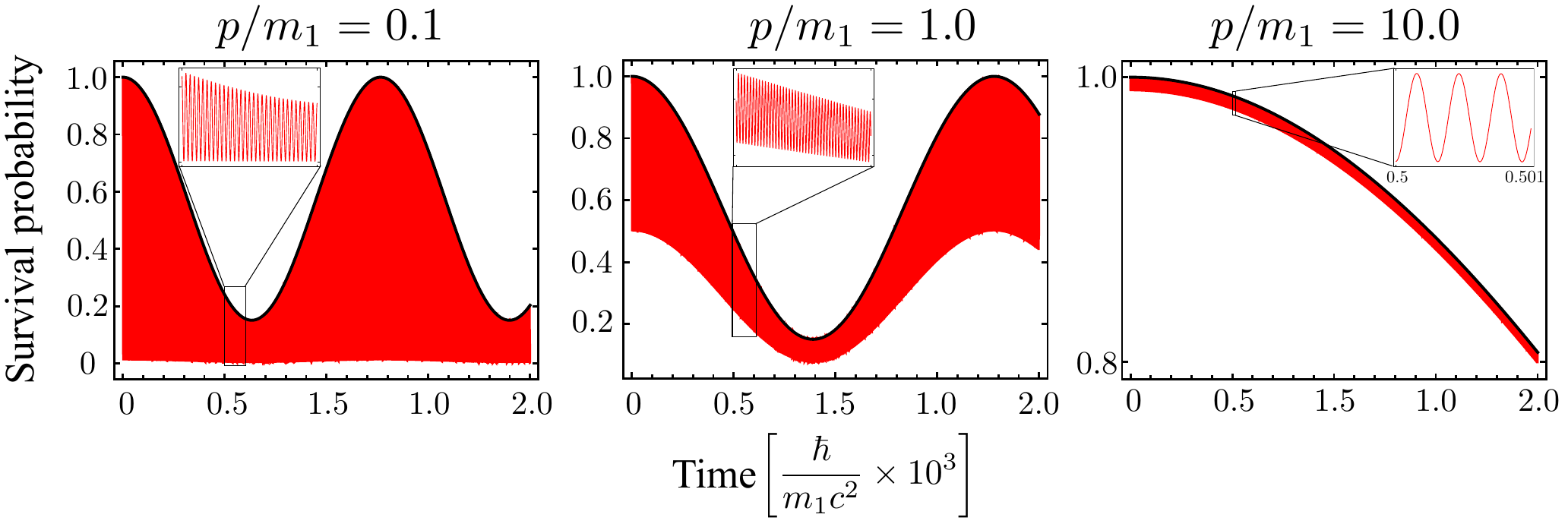}
\caption{Survival probability $\mathcal{P}_{e\rightarrow e}(t)$ as a function of time. The black curves indicate the standard survival probability formula \eqref{standard} and the red curves depict the full formula including the fast chiral oscillations (depicted in the insets). Parameters: $\sin^2\theta = 0.306$, $m_2^2 = \Delta_{21}^2 + m_1^2$, with $\Delta_{21}^2/m_1^2 = 0.01$.}
\label{Fig:03}
\end{figure}

We show the survival probability \eqref{survivProbE} in Figure \ref{Fig:03} for several values of $p/m$. As already anticipated in the previous section and in \cite{Bernardini:2004epjc,Bernardini:2005prd,Bernardini:2011fph}, the chiral oscillations induce just small corrections to the state's dynamics in the ultra-relativistic regime $p\gg m$. For the non-relativistic regime $p \lessapprox m_{1,2}$, the full oscillation formula differs significantly from the standard result due to high amplitude chiral oscillations. For the parameters of Figure \ref{Fig:03}, the mass difference is very small, and the flavor oscillations have a time scale much longer than the chiral oscillations.

\subsection{Chiral oscillations effect on the average flavor oscillation}

Given the different time scales of the flavor and chiral oscillations, we consider the averaged survival probability over one flavor oscillation, defined as
\begin{equation}
\bar{\mathcal{P}}_{e\rightarrow e} = \frac{1}{\tau_{21}} \displaystyle \int_{0} ^{\tau_{12}} \mathcal{P}_{e\rightarrow e}(t),
\end{equation}
where $\tau_{12} = \frac{4 \pi}{E_{p, m_2} - E_{p, m_1}}$ is the period of one flavor oscillation according with the standard formula \eqref{standard}. The time integration of \eqref{survivProbE} leads to
\begin{equation}
\begin{aligned}
\bar{\mathcal{P}}_{e\rightarrow e} &= 1 - \frac{\sin^2\left( 2 \theta \right)}{ 4}\left[2+ \frac{f_{21}}{ 4 \pi} \sin \left( \frac{4 \pi}{ f_{21}} \right) \left( \frac{p^2}{E_{p, m_2} E_{p, m_1}} -1\right) \right] - \frac{1}{2} \left( \frac{m_1^2}{ E_{p, m_1}^2} \cos^4 \theta + \frac{m_2^2}{ E_{p, m_2}^2} \sin^4 \theta \right)  \\
&+ \frac{f_{21}}{4 \pi(1-f_{21}^2)} \sin \left( \frac{4 \pi}{ f_{21}} \right) \left[ (1+f_{21})\frac{m_1^2}{ E_{p, m_1}^2} \cos^4 \theta + (1- f_{21})\frac{m_2^2}{ E_{p, m_2}^2} \sin^4 \theta  \right]
\end{aligned}
\label{fullavg}
\end{equation}
with $f_{21} =( E_{p, m_2} - E_{p, m_1} )/( E_{p, m_2} + E_{p, m_1} )$. Considering that the average of the standard oscillation formula \eqref{standard} is given by
\begin{equation}
\bar{\mathcal{P}}_{e \rightarrow e}^{S} = 1 - \frac{\sin^2\left( 2 \theta \right)}{2},
\end{equation}
one concludes that the quantity $\bar{\mathcal{P}}_{e \rightarrow e}^{S}-\bar{\mathcal{P}}_{e\rightarrow e}$ contains all the chiral oscillation effects on the averaged survival probability, and therefore properly quantifies the effects of chiral oscillations on flavor oscillations.

\begin{figure}[t]
\centering
\includegraphics[width = 13 cm]{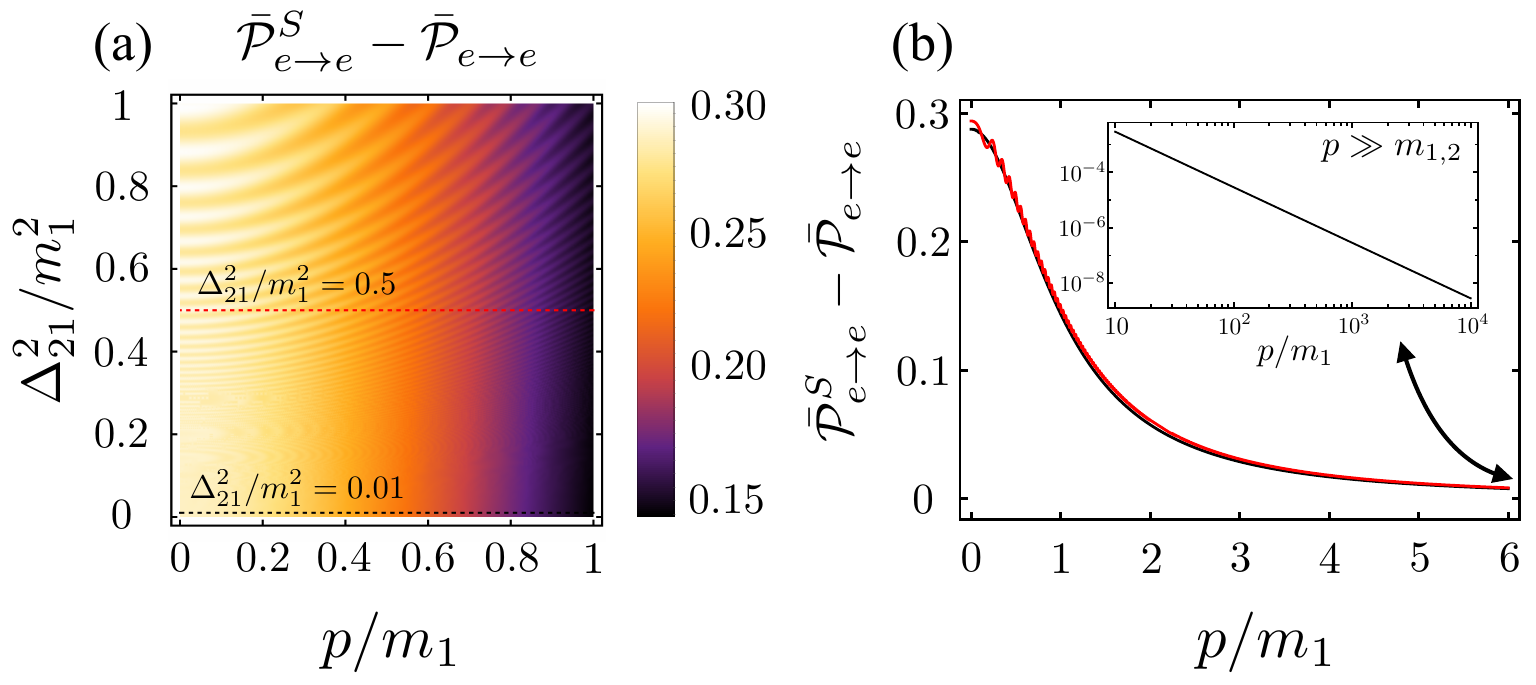}
\caption{Difference between the averaged standard survival probability $\bar{\mathcal{P}}_{e\rightarrow e}^{S}= 1 - \frac{\sin^2\left( 2 \theta \right)}{4}$ and the full survival probability including chiral oscillation effects \eqref{fullavg} (a) as a function of the state's momentum $p$ and of the squared mass difference $\Delta_{21}^2 = m_2^2 - m_1^2$ and (b) as a function of the momentum $p$ for $\Delta_{21}^2/m_1^2 = 0.01$ (black) and $0.5$ (red). The inset shows the chiral oscillations corrections for the ultra-relativistic regime as given by \eqref{UR} for $\Delta_{21}^2/m_1^2 = 0.01$ in logarithmic scale. Other parameter as in Fig.~\ref{Fig:03}.}
\label{Fig:04}
\end{figure}

In Figure \ref{Fig:04} we show the difference $\bar{\mathcal{P}}_{e \rightarrow e}^{S}-\bar{\mathcal{P}}_{e\rightarrow e}$ as a function of the neutrino momentum and the squared mass difference and as a function of $p/m_1$ for fixed values of the squared mass difference. The difference between the averaged standard probability and the full one becomes $\gtrsim 0.1$ for $p\lesssim m_1$, indicating the influence of chiral oscillations on flavor oscillations. In the non-relativistic regime, the difference between the probabilities can be as big as $\sim 40 \%$ for typical values of the mixing angle. Moreover, we notice that for bigger values of the mass difference, oscillations in the probability (with the momentum) are observed. This is due to the terms $\propto  f_{21}  \sin \left( 4 \pi/ f_{21} \right)$ in \eqref{fullavg}, which depend on the sum of the energies, and oscillate with the momentum when $m_1$ and $m_2$ are well separated. In the ultra-relativistic regime $p \gg m_{2,1}$, we can expand the full averaged formula \eqref{fullavg} with respect to $m_{2,1}/p$ as to have
\begin{equation}
\label{UR}
\bar{\mathcal{P}}_{e\rightarrow e}^{{\rm{UR}}} = \bar{\mathcal{P}}_{e\rightarrow e}(p \gg m_{2,1}) =\bar{\mathcal{P}}_{e \rightarrow e}^{S} + \frac{\Delta_{21}^2}{4 p^2} \cos(2 \theta) -\frac{\Sigma_{21}^2}{4 p^2} \left[1 - \frac{\sin^2(2 \theta)}{2} \right] + \mathcal{O}[m_{1,2}^4/p^4],
\end{equation}
The corrections to the standard result are thus proportional to both the squared mass difference $\Delta_{21}^2=m_2^2 - m_1^2$ and to the squared mass sum $\Sigma_{21}^2 = m_2^2 + m_1^2$. The inset of Fig.~\ref{Fig:04}(b) depicts the corrections $\bar{\mathcal{P}}_{e\rightarrow e}^{S}- \bar{\mathcal{P}}_{e\rightarrow e}^{{\rm{UR}}}$ in logarithmic scale.

\subsection{Chiral oscillations and tests for Dirac and Majorana neutrinos}

Throughout the main text, we have assumed that the initial neutrino state is a left-handed chiral eigenstate, as a consequence of the properties of weak interactions, in which neutrinos are created. We do not address here questions related to the production of the initial state (such as its localization features). 

In proposals for non-relativistic neutrinos detection \cite{Baracchini:2018wwj,Long:2014zva,Roulet:2018fyh}, it was found that the nature of neutrinos (Dirac or Majorana) influences directly the experiment's measurement rate. At low energies Dirac neutrinos can be detected as left-chiral and negative helicity, while Majorana neutrinos can also be detected as right-chiral and positive helicity \cite{Ge:2020}. When chiral oscillations are considered, the overall flux of detected neutrinos is halved due to chiral oscillations. As we show, Dirac neutrinos chiral oscillations exhibit a strong effect in the non-relativistic regime: oscillations from left-chiral to right-chiral become more prominent. In the case of Majorana neutrinos, the possible transition probabilities (cf. in the appendix) have the same energy dependence as in the Dirac case, and the survival/oscillation probabilities are equal to those for Dirac particles. Therefore, the measurement of non-relativistic neutrinos will unavoidably be affected by chiral oscillations and are a probe of those effects. Since both Majorana and Dirac neutrinos are subjected to chiral oscillations, in the standard scenario, chiral oscillations and their impact in other relevant quantities, such as absorption rates, can be computed once the nature of the neutrinos is known (see the discussion in \cite{Ge:2020}). 

The mass term for Majorana particles can be generalized such that left and right-handed components of the neutrino have different masses \cite{Cheng}. In the case of a vanishing Dirac mass term, but for different left and right Majorana mass terms, we recover the same survival probability and chiral oscillations behavior, as discussed in the appendix. In general case of mixed Dirac-Majorana masses, there will be interferences between the different chiral components, which will lead to further imprints in the survival probability. This most general case is also discussed in the appendix. Furthermore, in the presence of an external magnetic field, the coupled flavor-chiral oscillations are modified due to the non-minimal coupling to the field \cite{Bernardini:2006epjc02}. Majorana neutrinos exhibit different oscillations properties in the presence of a magnetic field, which can also be used as a way to distinguish Dirac from Majorana particles \cite{Dvornikov:2009,Dvornikov:2010}. Works within this framework have focused on relativistic neutrinos, an interesting question is whether such effects are more prominent in the non-relativistic regime.

As for the observability of such an effect we note that, in connection to the aforementioned proposal \cite{Baracchini:2018wwj,Long:2014zva,Roulet:2018fyh}, neutrinos from the C$\nu$B are expected to be extremely non-relativistic and therefore subject to chiral oscillation effects. The current temperature of the C$\nu$B $T_{{\rm{C}} \nu {\rm{B}}}$ is related to the temperature of the cosmic microwave background (CMB) $T_{\rm{CMB}}$ via $T_{{\rm{C}} \nu {\rm{B}}} \simeq (4/11)^{1/3}T_{\rm{CMB}} \sim 0.168$ meV. Since neutrinos maintain a Fermi-Dirac like distribution after decoupling \cite{Lesgourgues:2006}, one can obtain the root mean square momentum as $\bar{p}_0 \sim 0.603$ meV \cite{Long:2014zva}. Considering the lightest mass as $m_1 \sim 0.1$ eV, these values would give the maximum effect persisting for $m_1$ down to the meV order. A more complete recent study of the effects of chiral oscillations in the C$\nu$B can be found in \cite{Ge:2020}.

\section{Conclusions}

In this paper we have described dynamical features of chiral oscillations in the non-relativistic regime, both from a general perspective including their relation with \textit{Zitterbewegung}, to the study of flavor oscillations for non-relativistic neutrinos. Such framework is especially interesting for upcoming experiments that will be capable to measure the C$\nu$B \cite{Baracchini:2018wwj}, for which  chiral oscillations should have a prominent influence.

We considered the free plane wave propagation of an initial left-chiral state, as those generated, for example, via weak interactions. The left-right chiral oscillations are induced by the mass term of the Dirac equation, and we computed the characteristic oscillation length in the non-relativistic limit: in this regime, the survival probability of the initial state could vanish. For masses of the order of ${\rm{eV}}$, the chiral oscillation length is $\sim 10^{-6} {\rm{m}}$.

We then specialized our results to describe chiral oscillations in neutrino propagation by considering a two-flavor problem described within the Dirac bispinor framework. While, as discussed in \cite{Bernardini:2004epjc,Bernardini:2005prd,Nishi:2006prd,Bernardini:2011fph}, such corrections are negligible for ultra-relativistic particles, we verified that chiral oscillations are relevant in the non-relativistic dynamical regime. We obtained a modified flavor oscillation formula including the chiral oscillations effects, compatible with known results in the literature \cite{Nishi:2006prd}. While the standard flavor oscillation term goes with the mass eigenstate energy difference, chiral oscillations depend on the individual energies of the mass eigenstates.

To quantify the effects of chiral oscillations, we computed the averaged survival probability and observed that, for non-relativistic neutrinos, chiral oscillations can affect averaged flavor oscillations up to $\sim 40\%$ of the standard value. This is especially relevant for upcoming C$\nu$B tests which will probe non-relativistic neutrinos. Furthermore, Majorana neutrinos exhibit the same chiral oscillations properties of Dirac neutrinos, although the framework in the Majorana case is more enhanced. In particular, we show in the appendix that any bispinor that is its own charge-conjugate (that satisfies the Majorana equation), must have vanishing average chirality. Nevertheless, for a neutrino state initially in a left-handed configuration evolving under the Majorana equation, one gets a survival probability compatible with the Dirac case. A further investigations of those effects in the non-relativistic dynamical regime for a mixed Dirac-Majorana mass deserve a more detailed analysis and can shed a light on other possible distinctions to the usual scenario based on oscillation experiments (see for example \cite{Bilenky:1981}).

While our results were derived for two flavors, the extension of the formalism to three flavors is straightforward with qualitatively similar conclusions. Moreover, wave packets can be readily considered \cite{Bernardini:2005prd}. In these case, there would be additional effects which also have a non-trivial influence in flavor oscillation. Nevertheless, since chiral oscillations depend on the individual energies of the mass eigenstates, decoherence effects due to a wave-packet treatment, which depend on the difference of neutrino masses, do not influence chiral oscillations. In fact, the latter are intrinsic to the Dirac bispinors, and thus to each mass eigenstate. Finally we would like to point that the formalism adopted here is an effective description: flavor oscillations are correctly described in the framework of quantum field theory \cite{Blasone:1995aop,Blasone:1998hf}, to which we plan to extend our treatment in the future.

\textit{Acknowledgements}: 
We thank L. Smaldone for useful discussions. VASVB acknowledges financial support from the Max Planck Gesellschaft through an Independent Max Planck Research Group.

\section*{Appendix - Majorana bispinors and chiral oscillations}

The Lagrangian for a massive fermion field $\Psi$ is:
\begin{equation}
\mathcal{L}=  \bar{\Psi}\hat{\gamma}^\mu \hat{p}_\mu \Psi + \mathcal{L}_{\rm{mass}},
\end{equation}
where the mass term $\mathcal{L}_{\rm{mass}}$ depends on the nature of field: Dirac or Majorana. For Dirac fermions, the mass term reads
\begin{equation}
\mathcal{L}_{\rm{mass}}^{(D)}= - m_{\rm{D}} \bar{\Psi} \Psi,
\end{equation}
which can be rewritten in terms of the left and right-handed components of the field $\Psi_L = (1-\hat{\gamma}_5)\Psi/2$, $\Psi_R = (1+\hat{\gamma}_5)\Psi/2$ as
\begin{equation}
\label{DiracLagr}
\mathcal{L}_{\rm{mass}}^{(D)}= - m_{\rm{D}}(\bar{\Psi}_R \Psi_L +\bar{\Psi}_L \Psi_R ).
\end{equation}
The equation of motion for the total Lagrangian in this case is the Dirac equation whose solutions are given in terms of the Dirac bispinors presented in the main text. The mass term is responsible for coupling the left-handed to the right-handed components of the bispinor, yielding to chiral oscillations. Otherwise, the Majorana mass terms reads
\begin{equation}
\mathcal{L}^{(M)}=- m_{\rm{R}}( \bar{\Psi}_R^c \Psi_R + {\rm{h.c.}} ) - m_{\rm{L}}( \bar{\Psi}_L^c \Psi_L + \rm{h.c.} ),
\end{equation}
where $\Psi^c = i \hat{\beta} \hat{\alpha}_y \Psi^*$ denotes the charge-conjugated field and the left and right handed components have different masses.
The above Lagrangian term can be rewritten as \cite{Cheng}
\begin{equation}
\mathcal{L}^{(M)} = - m_{\rm{R}} \bar{\omega} \omega -m_{\rm{L}} \bar{\chi} \chi,
\end{equation}
where
\begin{equation}
\label{MajField01}
\begin{aligned}
\chi &= \Psi_L +\Psi_L^c, \\
\omega &= \Psi_R+\Psi_R^c. \
\end{aligned}
\end{equation}
Notice that $\chi^c = \chi$ and $\omega^c = \omega$. A more general mass term includes both Dirac an Majorana mass terms\footnote{For convenience we have included factors $2$ in the definition of the masses.}:
\begin{equation}
\label{GeneralMassTerm}
\mathcal{L}^{(DM)}=-\frac{m_{\rm{D}}}{2}(\bar{\Psi}_R \Psi_L +\bar{\Psi}^c_R \Psi^c_L + {\rm{h.c.}} ) - \frac{m_{\rm{R}}}{2}( \bar{\Psi}_R^c \Psi_R + \rm{h.c.} ) - \frac{m_{\rm{L}}}{2}( \bar{\Psi}_L^c \Psi_L + {\rm{h.c.}} ).
\end{equation}
While in the main text we have considered the ``usual'' Dirac equation (the one obtained via the mass term \eqref{DiracLagr}), we now consider briefly quantum oscillations under the Majorana mass term and under the general Majorana-Dirac mass term.

\subsection*{Degenerate Majorana mass term in the bispinor formalism}

While Majorana fermions are usually described within the framework of quantum field theory, we will consider from now own the framework of single particle relativistic quantum mechanics and, as in the main text, consider the dynamics of the bispinors. The single particle relativistic quantum mechanics of the Dirac equation with a Majorana mass term has its own peculiarities which we will not discuss here, see for example \cite{Hashimi:2017,Arodz:2019,Arodz2:2019,Dvornikov:2009}.

In this section we consider $m_{\rm{L}}= m_{\rm{R}}= m_{\rm{M}}$, for which the equation of motion reads
\begin{equation}
i \partial_t \Psi(\bold{x},t) = -i \hat{\bold{\alpha}} \cdot \nabla \Psi(\bold{x},t) + m_{\rm{M}} \hat{\beta}\Psi^{c}(\bold{x},t),
\end{equation}
or in the language used through the text
\begin{equation}
\label{MajEq}
i \partial_t \ket{\psi} = \hat{\bold{p}}\cdot \hat{\bold{\alpha}} \ket{\psi} + m_{\rm{M}} \hat{\beta}\ket{\psi}^{c},
\end{equation}
where $\ket{\psi}^{c}=i \hat{\beta} \hat{\alpha}_y\ket{\psi}^{*}$ is the charge-conjugated bispinor. For now own, we call the Dirac equation with a Majorana mass term as the Majorana equation. 

We notice that \eqref{MajEq} decouples into two independent sectors:
\begin{equation}
\label{RegDir}
i \partial_t \ket{\psi_\pm} = \hat{\bold{p}}\cdot \hat{\bold{\alpha}} \ket{\psi_\pm} \pm m_{\rm{M}} \hat{\beta}\ket{\psi_\pm},
\end{equation}
where
\begin{equation}
\ket{\psi_\pm}=\frac{\ket{\psi} \pm \ket{\psi}^{(c)}}{2}.
\end{equation}
The $\ket{\psi_\pm}$ are eigenstates of charge conjugation, and following the literature we call them Majorana bispinors. We have now the following scenario: a Dirac bispinor is an unconstrained bispinor with dynamical evolution given by the Dirac equation; a Majorana bispinor satisfies the Majorana condition (is an eigenstate of charge conjugation) AND has the dynamics given by the Dirac equation \eqref{RegDir}. It is common to discuss the properties of Majorana bispinors in the so-called Majorana representation of the Dirac matrices, where the Dirac matrices are all imaginary and charge conjugation is just a complex conjugation operation. To better compare Majorana and Dirac bispinors, we keep with the chiral representation of the Dirac matrices. The single particle relativistic quantum mechanics for Majorana particles is obtained by considering the sector related to $\ket{\psi_+}$. 

The Majorana condition imposes a constrain in the form of a bispinor. In particular it relates the upper to the lower (or, in our language the left to the right-handed components) of the bispinor. Let $\ket{\psi^{(M)}}$ be a Majorana bispinor satisfying the condition $\ket{\psi^{(M)}}^c = \ket{\psi^{(M)}}$. We can readily show that, if we write 
\begin{equation}
\label{GeneralMajorana}
\ket{\psi^{(M)}} =\begin{bmatrix}
  \ket{\xi_R^{(M)}} \\
   \ket{\xi_L^{(M)}}
\end{bmatrix},
\end{equation}
the charge conjugation condition implies that $\ket{\xi^{(M)}_L}=i \hat{\sigma}_y \ket{\xi_R^{(M)}}^*$ and a general bispinor satisfying the Majorana condition must be of the form
\begin{equation}
\label{RightMaj}
\ket{\psi^{(M)}_R} =\begin{bmatrix}
  \ket{\xi_R^{(M)}} \\
  i \hat{\sigma}_y \ket{\xi_R^{(M)}}^*
\end{bmatrix},
\end{equation} 
which is often called a right-handed Majorana bispinor, or of the form
\begin{equation}
\label{LeftMaj}
\ket{\psi^{(M)}_L} =\begin{bmatrix}
  - i \hat{\sigma}_y \ket{\xi_L^{(M)} }^* \\
  \ket{\xi_L^{(M)}}
\end{bmatrix},
\end{equation}
called a left-handed Majorana bispinor. Those bispinors correspond to \eqref{MajField01}, as can be seen by writing $\psi_{R,L}$ in a two component notation and using the explicitly form of the charge conjugation operator. We conclude that any bispinor satisfying the Majorana condition must have both left and right-handed components, this is a well known and commented through the literature (see e.g. \cite{Pal:2011}). 

The average chirality of a generic bispinor is given by
\begin{equation}
\langle \hat{\gamma}_5 \rangle = \langle \psi \vert \hat{\gamma}_5 \vert \psi \rangle.
\end{equation}
Since this is a real quantity, we have
\begin{equation}
 \langle \psi \vert \hat{\gamma}_5 \vert \psi \rangle= \left( \langle \psi \vert \hat{\gamma}_5 \vert \psi \rangle \right) = \left(\langle \psi \vert \right)^* \hat{\gamma}_5 \left(\vert \psi \rangle \right)^*,
\end{equation}
where in the last equality we have explicitly assumed that we are working in the chiral representation (thus $\hat{\gamma}_5$ is real). By recalling then that $\hat{\gamma}_y^2 = - \hat{I}_4$ we can write
\begin{equation}
\begin{aligned}
 \langle \psi \vert \hat{\gamma}_5 \vert \psi \rangle &=  \left(\langle \psi \vert \right)^* \hat{\gamma}_5 \left(\vert \psi \rangle \right)^* \\
 &= \left(\langle \psi \vert \right)^* \hat{\gamma}_5 i \hat{\gamma}_ y \left( i \hat{\gamma}_ y \left(\vert \psi \rangle \right)^* \right) \\
 &= - \left(\langle \psi \vert \right)^* i \hat{\gamma}_ y \hat{\gamma}_5 \vert \psi \rangle^c,
 \end{aligned}
\end{equation}
where in the last line we have used that $\{ \hat{\gamma}_y, \hat{\gamma}_y \} = 0$. Finally, since $ \langle \psi \vert^c = \left(\langle \psi \vert \right)^* i \hat{\gamma}_ y$ (because $\hat{\gamma}_y ^\dagger = - \hat{\gamma}_y$) we have
\begin{equation}
\label{avechirc}
\langle \hat{\gamma}_5 \rangle = \langle \psi \vert \hat{\gamma}_5 \vert \psi \rangle =  - \langle \psi \vert^c \hat{\gamma}_5 \vert \psi \rangle^c = - \langle \hat{\gamma}_5 \rangle_c,
\end{equation}
where we have defined $ \langle \hat{\gamma}_5 \rangle_c= \langle \psi \vert^c \hat{\gamma}_5 \vert \psi \rangle^c$. The above equations brings no new information: the average chirality of the charge conjugated spinor is opposite to the average chirality of the spinor. For example, if $\ket{\psi}$ is right-handed, then $\langle \hat{\gamma}_5 \rangle = 1$, and thus $\langle \hat{\gamma}_5 \rangle_c = -1$, in other words, $\ket{\psi}^c$ is a left-handed spinor. Charge conjugation flips the chirality of a spinor.

For a Majorana bispinor $\vert \psi \rangle^c = \vert \psi \rangle$, thus $\langle \hat{\gamma}_5 \rangle_c = \langle \hat{\gamma}_5 \rangle$ and from \eqref{avechirc} we have
\begin{equation}
\langle \hat{\gamma}_5 \rangle = - \langle \hat{\gamma}_5 \rangle \rightarrow \langle \hat{\gamma}_5 \rangle  = 0.
\end{equation}
Thus any bispinor that is its own charge conjugate must have zero average chirality. Notice that these calculations were performed for generic bispinors, and the only assumption is the Majorana condition. Therefore, if a bispinor satisfy the Majorana condition it has zero chirality at any time, independent of the specific time evolution it follows. This conclusion can also be reached by computing $\langle \hat{\gamma}_5 \rangle$ explicitly for a bispinor satisfying the Majorana condition, that's it, for a bispinor of the forms (\ref{RightMaj},\ref{LeftMaj}).

We now illustrate this point by considering the time evolution of a Majorana bispinor constructed with the Dirac bispinor of the main text. The Majorana bispinor $\ket{\psi^{(M)}}$  is thus written as
\begin{equation}
\ket{\psi^{(M)}} = \frac{ \ket{\psi^{(D)}} + \ket{\psi^{(D)}}^c }{\sqrt{2}}.
\end{equation}
In what follows we use the following relations for the bispinors  (see Eq. \eqref{DiracBispinors}):
\begin{equation}
\left(e^{i \bold{p} \cdot \bold{x}} \ket{u_\pm(p,m)} \right)^c =\pm e^{-i \bold{p} \cdot \bold{x}}\ket{v_\mp(p,m)},
\end{equation}
thus, for our calculations in the momentum space, the charge conjugation has the effect $$\ket{u_\pm(p,m)}^c= \pm \ket{v_\mp(-p,m)}.$$
Adopting the shorthand notation
\begin{equation}
\begin{aligned}
f_{p,m} &= 1 - \frac{p}{E_{p,m}+m},\\
g_{p,m} &= 1 + \frac{p}{E_{p,m}+m},\\
\mathcal{N}_{p,m} &= \sqrt{\frac{E_{p,m}+m}{4 E_{p,m}}},
\end{aligned}
\end{equation}
notice that $f_{-p,m} = g_{p,m}$. We can write the plane wave Dirac bispinor describing the evolution of a \textbf{Dirac} state initially in left-handed and negative helicity, see eq. \eqref{TempEvo}, as
\begin{equation}
 \ket{\psi_m(t)} \equiv  \ket{\psi^{(D)}_m(t)}  = \mathcal{N}_{p,m} \left[g_{p,m} e^{- i E_{p,m} t }\ket{u_-(p,m)} -  f_{p,m} e^{i E_{p,m} t} \ket{v_-(-p,m)} \right]. 
\end{equation}
The corresponding \textbf{Majorana} bispinor state is given by
\begin{equation}
\label{MajTime}
\ket{\psi^{(M)}_m(t)}  = \frac{\mathcal{N}_{p,m}}{\sqrt{2}} \left[ g_{p,m}  e^{- i E_{p,m} t }  \left(\ket{u_-(p,m)} - \ket{u_+(p,m)} \right) -  f_{p,m} e^{i E_{p,m} t} \left(  \ket{v_-(-p,m)} + \ket{v_+(-p,m)} \right) \right]. 
\end{equation}
Here we have used that the charge conjugation includes a complex conjugation, which has to be taken into account in the spatial dependence of the state, which can be grouped with the terms from $\ket{\psi^{(D)}_m(t)}$ by making $\bold{p}\rightarrow - \bold{p}$, in a similar fashion to what is done to group the negative-energy part of the wave function.

At $t= 0$ the state reads
\begin{equation}
\label{InitMaj}
\ket{\psi^{(M)}_m(0)} \equiv \ket{\psi^{(M)}(0)} = \frac{1}{\sqrt{2}}\begin{bmatrix}
    -\ket{+}  \\
    \ket{-}  
\end{bmatrix},
\end{equation}
while for the Dirac case we had
\begin{equation}
\ket{\psi^{(D)}_m(0)} \equiv \ket{\psi^{(D)}(0)} = \begin{bmatrix}
    0  \\
  \ket{-}  
\end{bmatrix}.
\end{equation}
Whether $\ket{\psi^{(D)}_m(0)}$ or $\ket{\psi^{(M)}_m(0)}$ really represents the initial state of a mass eigenstate depends on the specific weak processes that generates the particle, and will not be discussed here.

The survival probability of \eqref{InitMaj} is given by
\begin{equation}
\mathcal{P}^{(M)}_{S}(t) = \left \vert \inn{ \psi^{(M)}_m(0)}{\psi^{(M)} _m (t)} \right \vert^2=1 - \frac{m^2}{E_{p,m}^2}\sin^2(E_{p,m} t) = \mathcal{P}(t).
\end{equation}
Thats it, the survival probability of the considered initial state is the same as in the Dirac case. This is consistent with the conclusions drawn in \cite{Ge:2020} regarding chiral oscillations: The Majorana neutrino also exhibit quantum oscillations. In this case, the upper right-handed and positive helicity component can oscillate to a left-handed positive helicity, while the lower left-handed and negative helicity oscillates to a right-handed negative helicity. This two oscillation channels yield the survival probability obtained in the above equation. This can be further understood by considering the probability $\mathcal{P}^{(M)}_{L,-}$ of measuring $[\, 0, \, 0, \, 0, 1 ]^T$ (a left handed and negative helicity state), which is given by
\begin{equation}
\mathcal{P}^{(M)}_{L,-}=\frac{\mathcal{P}(t)}{2}.
\end{equation}
This probability is half of the survival probability for the Dirac bispinor state \eqref{chiralosc01}. Since \eqref{MajTime} is a superposition of both positive and negative helicities, the probability that the state is in a right-handed and positive helicity is not null. In fact, such probability is given by
 \begin{equation}
 \label{RightOsc}
\mathcal{P}^{(M)}_{R,+} (t) = \frac{\mathcal{P}(t)}{2}. 
\end{equation}
We therefore notice that
\begin{equation}
\label{MajProbs}
\mathcal{P}^{(M)}_{R,+} (t) + \mathcal{P}^{(M)}_{L,-} (t) = \mathcal{P}(t).
\end{equation}
Finally, the average chirality of \eqref{MajTime} vanishes: as discussed in the beginning of this section, the Majorana condition implies that the bispinor is a superposition of left and right chirality components, thus its average chirality vanishes.

\subsection*{Quantum oscillations of left-handed chiral eigenstates}

The calculations above are an illustrative example of the dynamics of a Majorana bispinor obtained with the corresponding Dirac bispinors. We can instead turn our attention to the formalism of \cite{Esposito:1997,Esposito:1998} to evaluate the time evolution of an initially left-handed state.We generalize the result in the aforementioned references and consider through this section the general Dirac-Majorana mass term \eqref{GeneralMassTerm}, which can be write in terms of the bispinors $\chi=\Psi_L + \Psi_L^c$ and $\omega=\Psi_R + \Psi_R^c$ as
\begin{equation}
\mathcal{L}^{(DM)}=-\frac{m_{\rm{D}}}{2}(\bar{\omega} \chi + \bar{\chi} \omega) - \frac{m_{\rm{R}}}{2} \bar{\omega} \omega - \frac{m_{\rm{L}}}{2} \bar{\chi} \chi.
\end{equation}
In other words, considering
\begin{equation}
\Psi = \Psi_L + \Psi_R = \begin{bmatrix}
  \psi_R\\
  \psi_L
\end{bmatrix},
\end{equation}
we are interested in the dynamics of $\vert \psi_L (t) \rangle$ under the general mass term \eqref{GeneralMassTerm}\footnote{Notice that the connection with the notation used in eq. \eqref{GeneralMassTerm} is:
\begin{equation}
\begin{aligned}
\Psi_L &=\begin{bmatrix}
  0\\
  \psi_L
\end{bmatrix},  \quad
\Psi_R =\begin{bmatrix}
  \psi_R\\
  0
\end{bmatrix}, \\
\Psi_L^c &=\begin{bmatrix}
  0\\
  \psi_L^c
\end{bmatrix},  \quad
\Psi_R^c =\begin{bmatrix}
  \psi_R^c\\
  0
\end{bmatrix}.
\end{aligned}
\end{equation}}.

This Lagrangian can be rewritten as
\begin{equation}
\mathcal{L}^{(DM)}=- m_1 \bar{\eta}_1 \eta_1 - m_2 \bar{\eta}_2 \eta_2,
\end{equation}
where \cite{Cheng}
\begin{equation}
\label{EigenEta}
\begin{aligned}
m_{1,2} &= \frac{m_{\rm{L}}+m_{\rm{R}} \pm \sqrt{(m_{\rm{L}}+m_{\rm{R}} )^2 + 4 m_{\rm{D}}^2}}{4}, \\
\eta_1 &= \cos{(\phi)} \chi - \sin{(\phi)} \omega, \\
\eta_2 &= \sin{(\phi)} \chi + \cos{(\phi)} \omega, \\
\tan{(2\phi)} &= \frac{2 m_{\rm{D}}}{m_{\rm{L}} -m_{\rm{R}}}.
\end{aligned}
\end{equation}

We then write $\eta_i$ ($i=1,2$) in terms of their right and left-chiral components
\begin{equation}
\eta_i =\begin{bmatrix}
  \eta_{i,R}\\
  \eta_{i,L}
\end{bmatrix},
\end{equation}
and considering negative helicity eigenstates, we get from the equations of motion:
\begin{equation}
\begin{bmatrix}
  E_{p,m_i} - p & -m_i\\
  -m_i & E_{p,m_i} + p
\end{bmatrix} \begin{bmatrix}
  \eta_{i,R}\\
 \eta_{i,L}
\end{bmatrix}=0,
\end{equation}
which gives the eigenenergies $E_{p,m_i}^{(\pm)} = \pm E_{p,m_i} = \pm \sqrt{p^2 + m_i ^2}$ and the eigenvectors
\begin{equation}
\begin{aligned}
\begin{bmatrix}
  f_{i}^{(+)}\\
f_{i}^{(-)}
\end{bmatrix} &= \begin{bmatrix}
  \cos{(\theta_i)} & \sin{(\theta_i)}\\
-\sin{(\theta_i)} & \cos{(\theta_i)}
\end{bmatrix}\begin{bmatrix}
  \eta_{i,R}\\
 \eta_{i,L}
\end{bmatrix}, \\
\tan{(2 \theta_i)} &= \frac{m_i}{p}.
\end{aligned}
\end{equation}
The eigenstates of the Hamiltonian time evolution given by
\begin{equation}
\vert f_{i}^{(\pm)} (t) \rangle = e^{\mp i E_{p,m_i} t} \vert f_{i}^{(\pm)} \rangle.
\end{equation}
To recover the dynamics of $\vert \psi_L (t) \rangle$, we backtrack the problem by using the relations \eqref{EigenEta} and then Eqs. \eqref{MajField01}, such that
\begin{equation}
\begin{aligned}
\vert \psi_L (t) \rangle &= \cos{(\phi)} \left[ \sin{( \theta_1)} e^{-i E_{p,m_1} t} \vert f_1^{+} \rangle + \cos{(\theta_1)} e^{i E_{p, m_1} t} \vert f_1^{-} \rangle \right] \\
&\, +\sin{(\phi)} \left[ \sin{(\theta_2)} e^{-i E_{p,m_2} t} \vert f_2^{+} \rangle + \cos{(\theta_2)} e^{i E_{p, m_2} t} \vert f_2^{-} \rangle \right],
\end{aligned}
\end{equation}
and therefore
\begin{equation}
\label{GeneralTempEvo}
\vert \psi_L (t) \rangle  = \mathcal{A}_L(t) \vert \psi_L \rangle + \mathcal{A}_R(t) \vert \psi_R \rangle + \mathcal{A}_L^c(t) \vert \psi_L^c \rangle + \mathcal{A}_R^c(t) \vert \psi_R^c \rangle. 
\end{equation}
The time dependent coefficients are given explicitly by
\begin{equation}
\begin{aligned}
\mathcal{A}_L(t) &=\cos^2 (\phi ) \left[\cos{(E_{p , m_1} t)} + i \cos{(2 \theta_1)}\sin{(E_{p , m_1} t)}\right]  \\
  &\quad +\sin^2 (\phi ) \left[\cos{(E_{p , m_2} t)} + i \cos{(2 \theta_2)}\sin{(E_{p , m_2} t)}\right], \\
\mathcal{A}_R(t) &= \frac{i \sin{(2 \phi )}}{2} \left[ \sin{(2\theta_1)}  \sin{(E_{p , m_1} t)} -  \sin{(2\theta_2)}  \sin{(E_{p , m_2} t)}\right], \\
\mathcal{A}_L^c(t) &= -\frac{1}{2}\sin (2 \phi ) \left[  \cos(E_{p , m_1} t) + i \cos(2 \theta_1) \sin(E_{p , m_1} t)   \right] \\
&\quad +\frac{1}{2}\sin (2 \phi ) \left[ \cos(E_{p , m_2} t) + i \cos(2 \theta_2) \sin(E_{p , m_2} t) \right], \\
\mathcal{A}_R^c(t) &= -i \left[ \cos^2 (\phi ) \sin(2 \theta_1) \sin{(E_{p , m_1} t)} + \sin^2 (\phi ) \sin(2 \theta_2) \sin{(E_{p , m_2} t)}\right].
\end{aligned}
\end{equation}
The survival probability of the state is given by
\begin{equation}
\mathcal{P}_S^{(G)} (t) = \vert \mathcal{A}_L(t) \vert^2
\end{equation}
while the average chirality is calculated by using that $\langle \psi_L \vert \hat{\gamma}_ 5 \vert \psi_L \rangle = \langle \psi_L^c \vert \hat{\gamma}_ 5 \vert \psi_L^c \rangle = -1$ and $\langle \psi_R \vert \hat{\gamma}_ 5 \vert \psi_R \rangle = \langle \psi_R^c \vert \hat{\gamma}_ 5 \vert \psi_R^c \rangle = 1$, such that
\begin{equation}
\langle \hat{\gamma}_5 (t) \rangle = \langle \psi_L (t) \vert \hat{\gamma}_5 \vert \psi_L (t) \rangle = -\vert \mathcal{A}_L (t) \vert^2- \vert \mathcal{A}_L^c (t) \vert^2 +\vert \mathcal{A}_R (t) \vert^2 + \vert \mathcal{A}_R^c (t) \vert^2.
\end{equation}

For a pure Dirac mass term, $m_{\rm{R}} = m_{\rm{L}} = 0$, the explicit temporal evolution \eqref{GeneralTempEvo} simplifies to
\begin{equation}
\ket{\psi_L^{({\rm{D}})} (t)} = \mathbb{A}(t) \ket{\psi_L } + \mathbb{B}(t) \ket{\psi_R },
\end{equation}
where 
\begin{equation}
\label{CoeffTime}
\begin{aligned}
\mathbb{A}(t) &= \cos{(E_{p,m_{\rm{D}}} t)} + i \frac{p}{E_{p,m}} \sin(E_{p,m_{\rm{D}}} t), \\
\mathbb{B}(t) &=-i \frac{m_{\rm{D}}}{E_{p,m}} \sin(E_{p,m_{\rm{D}}} t).
\end{aligned}
\end{equation}
This dynamics is in complete agreement with the one derived in the main text within the bispinor formalism. For a pure Majorana mass term with degenerate masses, $m_{\rm{R}} = m_{\rm{L}} = m_{\rm{M}}$ , and $m_{\rm{D}} = 0$, we have
\begin{equation}
\ket{\psi_L^{({\rm{M}})} (t)} = \mathbb{A}^{({\rm{M}})}(t) \ket{\psi_L } + \mathbb{B}^{({\rm{M}})}(t) \ket{\psi_R^c },
\end{equation}
where $\mathbb{A}^{({\rm{M}})}(t)$ and $\mathbb{B}^{({\rm{M}})}(t)$ are given by \eqref{CoeffTime} with the substitution $m_{\rm{D}} \rightarrow m_{\rm{M}}$. The survival probability of $\ket{\psi_L^{(\rm{M})} (t)}$ is the same as the one obtained for the state $\ket{\psi_L^{(\rm{D})} (t)}$, but the transition probability it to $\ket{\psi_R^c }$. Moreover, we conclude that $\langle \psi_L^{(\rm{D})} (t) \vert \hat{\gamma}_5 \vert \psi_L^{(\rm{D})} (t)\rangle = \langle \psi_L^{(\rm{M})} (t) \vert \hat{\gamma}_5 \vert \psi_L^{(\rm{M})} (t)\rangle$, thus chiral oscillations of the states $\ket{\psi_L^{({\rm{M}})} (t)}$ and $\ket{\psi_L^{({\rm{D}})} (t)}$ are equal. Those dynamical features are also the same in the more general case of $m_{\rm{D}}= 0$ and $m_{\rm{R}} \neq m_{\rm{L}}$. This is expected: for a vanishing Dirac mass, the equations of motion are fully decoupled in terms of the bispinor $\chi$ and $\omega$, and the dynamics of a state initially in a left-handed chirality is determined entirely by the sector related to $\chi$, which only induces oscillations between $\vert \psi_L \rangle$ and $\vert \psi_R^c \rangle$. In the case $m_{\rm{D}}= m_{\rm{L}}=0$, there are no quantum oscillations.

When both Dirac and Majorana masses are present, the survival probability and the average chirality display a more intricate dynamics. In this case, there will be interferences due to the different values of the masses $m_{1,2}$ which drive different oscillation channels. Further investigation of such effects will be addressed in a future work.

\end{document}